\documentclass[sigconf]{acmart}
\usepackage{graphicx} 
\usepackage{xcolor}
\usepackage{url}
\usepackage[normalem]{ulem}

\newif\ifdraft
\drafttrue 

\ifdraft
      
         \def\AR#1{{\color{blue}#1}}
         \def\SP#1{{\color{blue} #1}}
         \def\SPc#1{{\color{orange} #1}}
         \def\ARc#1{{\color{violet} #1}}

\else
        \def\AR#1{}
        \def\SP#1{}
        \def\SPc#1{}
        \def\ARc#1{}
\fi

\copyrightyear{2026}
\acmYear{2026}
\setcopyright{cc}
\setcctype{by}
\acmConference[CHI '26]{Proceedings of the 2026 CHI Conference on Human Factors in Computing Systems}{April 13--17, 2026}{Barcelona, Spain}
\acmBooktitle{Proceedings of the 2026 CHI Conference on Human Factors in Computing Systems (CHI '26), April 13--17, 2026, Barcelona, Spain}
\acmDOI{10.1145/3772318.3790832}
\acmISBN{979-8-4007-2278-3/2026/04}

\begin{document}




\title[Regulatory Practitioner Views on Automated Detection of Deceptive Design Patterns]{``What I'm Interested in is Something that Violates the Law'': Regulatory Practitioner Views on Automated Detection of Deceptive Design Patterns}

\author{Arianna Rossi}
\email{arianna.rossi@santannapisa.it}
\orcid{0000-0002-4199-5898}
\affiliation{%
  \institution{LIDER-Lab, DIRPOLIS, Sant'Anna School of Advanced Studies}
  \city{Pisa}
  \country{Italy}
}

\author{Simon Parkin}
\email{s.e.parkin@tudelft.nl}
\orcid{0000-0002-6667-0440}
\affiliation{%
  \institution{TU Delft}
  \city{Delft}
  \country{The Netherlands}
}

\begin{abstract}
Although deceptive design patterns are subject to growing regulatory oversight, enforcement races to keep up with the scale of the problem. One promising solution is automated detection tools,  many of which are developed within academia. We interviewed nine experienced practitioners working within or alongside regulatory bodies to understand their work against deceptive design patterns, including the use of supporting tools and the prospect of automation. Computing technologies have their place in regulatory practice, but not as envisioned in research. For example, investigations require utmost transparency and accountability in all the activities we identify as accompanying dark pattern detection, which many existing tools cannot provide. Moreover, tools need to map interfaces to legal violations to be of use. We thus recommend conducting user requirement research to maximize research impact, supporting ancillary activities beyond detection, and establishing practical tech adoption pathways that account for the needs of both scientific and regulatory activities.
\end{abstract}



\maketitle

\section{Introduction}

Deceptive design patterns\footnote{In this article, we use this term interchangeably with dark patterns, deceptive patterns or DPs.} (DPs) persist online. These are interface designs that manipulate users, to the extent that they are misled or forced into choices that provide added benefits to the service owner \cite{mathur2021makes}. This can usually include disadvantages to the user, such as being unable to reject unfair conditions, being signed up for subscriptions without a way to opt out, being nudged into accepting privacy-invasive options, or being misled into believing that products are about to become unavailable, among other issues.
As it is difficult for users to escape these situations or readily find better alternatives, these unfair business practices are increasingly being brought under legal scrutiny worldwide, with attention coming from policy-makers \cite{OECD_2022}, regulatory bodies \cite{EDPB_2023, BEUC_2022, GPEN_2024, ICPEN_2024} and legal scholars \cite{leiser2023dark, santos2025no, ISOLA2025106169}. 

Regulatory oversight does not by itself change the scale of the problem. Automating DP detection is seen as a promising way to keep pace with these deceptive patterns as they emerge online. This potential has been harnessed in a growing body of research studies, adding to a range of tools claiming to inform regulators in their efforts to manage the problem. In the articles, scientists hold that their models can assist regulators in their compliance assessments \cite{mathur2019dark, Nouwens2020, bouhoula2023automated, shi202550, Deceptilens_2025} and raise awareness among developers and designers, enabling them to avoid the implementation of deceptive designs \cite{chen2023unveiling}. This growth is happening in parallel to the emergence of Enforcement Technologies (EnfTech) \cite{Riefa_Coll_2024}, the use of computing technologies in regulatory activities.

In this interview study, we engage practitioners active in and around the regulatory sphere in the European Union, to answer the following research questions relating to deceptive patterns: 
\begin{enumerate}
    \item RQ1: In which activities do practitioners of the regulatory sphere employ computing technology? 
    \item RQ2: What requirements must computing technology these tools have to be used in their work?
    \item RQ3: In which enforcement activities are AI-based tools developed by academia most useful, and what requirements do they embed?
\end{enumerate}

We build on prior calls to mobilize research and regulatory action against deceptive patterns \cite{gray2024mobilizing}. With many options for tooling capabilities implied in academic research, and complementary research on catalogues and taxonomies of `dark', `fair', and `bright' patterns emerging \cite{mathur2019dark, gray2024ontology, mansur2023aidui, chen2023unveiling, Nie2024shadows, shi202550}, we may assume that enforcement is a pattern-matching problem. Much research capital is being invested in automated detection, which would benefit from an understanding of where -- and to what extent -- this effort can support intended practitioner-users. In this study, we explore to what extent the capacity and capabilities of regulatory practice are able to match the implied capacity of automated tooling, examining how automated DP detection and technology-supported enforcement collide, and what may emerge. 

To this end, we interviewed nine experienced professionals in and around regulation in various parts of the EU, covering data protection and consumer protection. This included representatives of regulators, policy-makers, business owners and civil society organizations that develop and/or deploy EnfTech for DP detection, therefore working across software development,  data science and law. The analysis of the interviews produced a mapping of regulatory activities, driving most strongly the reality that enforcement needs solid evidence of non-compliance against legal provisions (rather than research-centred catalogues of DPs), and that regulators need to choose which (comparatively few) cases they pursue from across the whole internet, a bottleneck that needs to be considered. Beyond detection, we find that there are many tasks of enforcement activities that automated tooling can help make more efficient, such as with traceable data collection and the identification of deceptive behaviours not immediately visible on the interface. 

Academic research tools can inspire capabilities to be developed further, either in-house or in collaboration with external partners, including research institutions and spin-offs. Our interviewees welcomed this kind of teamwork, but their utmost need for confidentiality must be carefully balanced with the openness of scientific research. We also recommend ways to align supply and demand so that the laudable work of academia can better meet the requirements of enforcement, for example, by explicitly mapping DP detection to legal violations, as well as by engaging early and meaningfully with prospective users of the tools. What is lacking are shared spaces where different stakeholders can carry out common experimentation, and practical pathways to streamline the adoption of existing technology, for which we offer some suggestions, including open-source tool development. All these activities would contribute to achieving digital fairness and augment the societal impact of scientific research.

In the remainder of the paper, we cover Related Work (Section \ref{sec:sota}), our Methodology (Section \ref{sec:method}), Results from the interviews (Section \ref{sec:results}) answering RQ1 and RQ2, and a reflection on the Results against current research capabilities (Section \ref{sec:reflect-tools}) answering RQ3. These are followed by Discussion and recommendations (Section \ref{sec:disc}), and a Conclusion to the paper (Section \ref{sec:conc}).

\section{Related Work}
\label{sec:sota}
In this state of the art we consider related research, from defining dark patterns, the tools that may be used to detect them, and how this aligns with regulations and regulatory practitioners.

\paragraph{Defining and categorizing DPs.}
The design community that has tried to counter the proliferation of dark patterns has first produced a plethora of lists and categorizations of dark patterns, some depending on the type of application under scrutiny (e.g., videogames \cite{zagal2013dark}, e-commerce \cite{mathur2019dark}) interface (e.g., voice interfaces \cite{Owens2022}), or device (e.g., home IoT devices \cite{Kowalczyk2023}). Significant scientific efforts have focused on the identification of DPs on cookie consent with a broad range of methodologies, e.g., \cite{Nouwens2020, Berens_2024, Kocyigit2024systematic}.

All of these efforts to define this burgeoning phenomenon resulted in a highly diversified, at times messy, vocabulary. Some design scholars have found the need to identify the constitutive features of these design patterns that are unlike others (in other words "what makes a dark pattern dark?" \cite{mathur2021makes}). Recently, numerous efforts have been made to unify this divergent research into coherent taxonomies, and therefore obtain a common language to refer to an emerging phenomenon across domains and professions \cite{mathur2019dark, gray2024ontology, Potel2023fair}. Some of these efforts \cite{mansur2023aidui, chen2023unveiling, Nie2024shadows, shi202550} were intentionally made to create a unified categorization to enable supervised machine learning to detect DPs automatically within datasets of interfaces.  Resorting to such methods implies that the task of recognizing DPs is a pattern-matching problem and that tools can be built for automating its solution.
\paragraph{Automated DP detection}
While some previous research relies on manual identification of DPs on websites or on previously scraped interfaces (e.g., \cite{Mildner2023social, Santos_purpose2021}), many authors now leverage Machine Learning (ML) techniques and state-of-the-art Large Language Models (LLMs). 
Mathur et al. \cite{mathur2019dark} pioneered large-scale crawling of e-commerce sites to identify DPs. Their work highlighted that some DPs may be unlawful, and thus their work can also benefit regulators and civil society,  and revealed the existence of third-party providers offering infrastructure for DPs (so-called dark-patterns-as-a-service).

Mansur et al. \cite{mansur2023aidui} introduced AidUI, a machine learning pipeline combining computer vision and NLP to classify deceptive patterns. Aimed at helping developers avoid DPs, it uses a rule-based framework and a custom-labelled dataset. With a similar approach, UIGuard \cite{chen2023unveiling} targets end-users and developers. The authors underscore that, when paired with crawlers, it could support monitoring and enforcement by regulators.
Nie et al. \cite{Nie2024shadows} benchmarked existing approaches, noticing that there are limitations in their classification capabilities, due to limited coverage and incomplete datasets.

Lately, there is an increasing amount of research on Large Language Models  for deceptive design detection (e.g., \cite{schafer2025don}). DeceptiLens \cite{Deceptilens_2025} and DPGuard \cite{shi202550} both leverage the multimodal capacities of LLMs to recognize a broad range of DPs in ad hoc created datasets. The first foresees DP researchers as the primary users, but also envisions the use for enforcement agencies. Lastly, other automated approaches explicitly target end-users, for example, by highlighting DPs found on the interface to strengthen their awareness \cite{fighting_2024,lu2024awareness}.

\paragraph{Deceptive design and the law}
Dark patterns have gained regulatory traction, with regulatory bodies like the European Data Protection Board (EDPB) \cite{EDPB_2023} and the European Organization for Consumer Protection (BEUC) \cite{BEUC_2022}, as well as legal scholars \cite{leiser2023dark, santos2025no, ISOLA2025106169}, actively mapping DPs to legal violations—often through interdisciplinary collaboration \cite{Santos_purpose2021, Gray_2021interaction}. In 2024, consumer protection agencies \cite{ICPEN_2024} and data protection authorities \cite{GPEN_2024} manually conducted sweep campaigns worldwide to assess the prevalence of this phenomenon on websites and inform regulatory action. This momentum is reflected in the EU Digital Strategy legislation (e.g., the Data Act, the Digital Services Act, the Digital Markets Act, the AI Act, and the Data Governance Act), which prohibits deceptive design practices that undermine user autonomy.

Researchers have identified opportunities for collaboration between academic and regulatory communities to enhance detection and enforcement \cite{gray2024ontology
, Gunawan_2025}. The terminology of deceptive design patterns now appears in formal sanctions \cite{Santos_Rossi_2023}, reinforcing the link between scholarship and enforcement. Public “halls of shame” further support accountability by documenting DPs \cite{DP-tipline}, and offer employees leverage to explain violations and resulting fines to employers \cite{brignull-2023}.

\paragraph{Automated detection of legal violations}
Due to the growing regulatory relevance of DPs, academic approaches increasingly link DP detection to compliance checking. Given the scale of digital services, manual review is impractical \cite{shi202550}, prompting calls for automation "to expedite discovery and enforcement" \citep[p. 10]{nouwens2022consent}. 

Deceptive design in cookie banners has received particular attention \cite{gundelach2023cookiescanner}, though their assessment may be felt as subjective and reliant on human judgment \cite{kirkman2023darkdialogs}. This is why tools like DarkDialogs \cite{kirkman2023darkdialogs} and Bouhoula et al.’s compliance measurement system \cite{bouhoula2023automated} do not only focus on DP detection, but also support legal analysis to determine whether violations occur. These tools are increasingly designed to assist regulators and legal scholars in evaluating compliance over time.

\paragraph{Actors in the detection and enforcement activities}
Despite promising academic developments, most of these tools lack engagement with prospective users and do not incorporate reflections on user requirements—except for \cite{Deceptilens_2025}, which involved DP experts, primarily researchers. Academic researchers often assume  utility for regulators and other stakeholders but have not validated these assumptions through direct engagement. This gap is significant, as DP detection and regulation activities face challenges, not only due to the scale of digital services but also due to definitional ambiguity. As noted, "detecting deceptive patterns is a difficult task with a moving target" \cite[p.2]{schafer2025don}, and some DPs exist in gray areas, making it hard to determine deception \cite{mathur2019dark}. 

Many regulations concerning personal data, consumer contracts, and digital services contain provisions that indicate lawful ways to design user interfaces, for example the fact that consent must be an "unambiguous indication" of the user agreement according to art. 4(11) General Data Protection Regulation (GDPR). Case law often provides interpretation of legal provisions, such as when the judges of the Court of Justice of the EU, in a landmark case \cite{planet49}, clarified that "unambiguous indication" means that consent boxes on a GUI cannot be preselected.  When those provisions are not respected, for example due to the use of a DP (e.g., privacy-invasive defaults \cite{EDPB_2023} are a typical example of a dark pattern), an interface may be considered unlawful. Regulatory bodies, such as data protection and consumer protection agencies, can initiate investigations and issue administrative fines to organizations that violate the law, even based on complaints from individuals. Given the sheer number of digital transactions, regulators show growing interest in automation beyond DP detection. A 2023 report on Enforcement Technologies (EnfTech) \cite{Riefa_Coll_2024}, involving global consumer protection agencies, reflects efforts to integrate computing technologies into consumer protection enforcement workflows.

Compliance monitoring also involves civil society organizations, in certain jurisdictions, which may accommodate complaints from individuals and submit mass complaints. While interviews with designers reveal tensions around knowingly implementing DPs \cite{zhang2024navigating}, interviews with enforcement practitioners are limited. Studies with UK case workers \cite{ceross2018examining} and EU professionals \cite{cervi2022and} highlight the need for a deeper understanding of enforcement processes and concerns about regulatory capacity.

\paragraph{Latent challenges} 
While several tools have been proposed by academia to support DP detection, questions remain regarding their practical relevance: do these tools risk over-promising? Do they adequately respond to user requirements? Have those requirements been systematically investigated? This work critically examines the disconnect between academic approaches that predominantly conceptualize DP detection as a pattern-matching problem, and the regulatory domain, which demands nuanced interpretation, legal reasoning, and contextual sensitivity.

\section{Methodology}
\label{sec:method}
In the following, we describe our study procedure, recruitment and participants, how we analysed the interview data and contrasted it against existing tools, as well as ethical considerations and limitations.

\subsection{Interview study design}
Our interview protocol involved many parts, as described in Appendix \ref{app:questions}. First, we asked about the participant's activities and role (relating to the legal domain), given that regulatory practitioners are rarely studied directly, especially in the domain of dark patterns. We then asked what the participant considered to be a legal violation, to relate this to existing dark patterns concepts. From these foundational questions, we moved to questions about the supporting tools that participants use in their role and why, and related requirements for using those tools. This focused on how tools are used for detection and enforcement. 

After this, we presented a short overview of three different types of enforcement tools, varying in their levels of automation. The first was `wall of shame'-style tools, which anyone can contribute to; the second was semi-automated tools using Machine Learning, which search websites based on fixed rules; the third was tools based on Large Language Models (LLMs), which in essence adapt rules to find variations of dark patterns. For each type of tool, we summarized their function, and their goal. This overview was used to act as a shared artifact during the interview, against which to discuss available technologies and their effectiveness; we are mindful that many tools exist in practice, with varying capabilities. Given that this is a technology that is developing at a fast rate -- what researchers may refer to as an `AI tool' may not be the same as what a practitioner has encountered as an `AI tool'. For that reason, we linked specific functionality and goals to each kind of tool. We encouraged participants to discuss the potential for such tools, referring to the content of the presentation so that we could ground comments in tangible features. The overview of the three types of tools used to support this conversation can be found in Appendix \ref{app:overview}.

After the tool presentation and associated discussion, we closed the interviews with discussion of automated tooling in the context of wider tooling developments within detection and enforcement (as `Enf-Tech' more broadly). At this point, we debriefed the participant as to our goal to explore the usefulness of automated tooling from research and the advancements it would represent.

\subsection{Recruitment and participants}
We recruited nine practitioners from within the European Union, reflecting our research focus, our familiarity with EU regulatory frameworks, and the prominence of deceptive patterns in EU regulations. Each participant brought extensive experience in the regulatory domain, working across jurisdictions and roles within or alongside regulatory bodies. For anonymity, we do not disclose their country, but Table \ref{tab:participants} details their organizational affiliations and declared areas of expertise. Many held interdisciplinary roles and all had direct experience with tools developed or used in their institutions.

Recruitment combined personal contacts and snowball sampling, with participants recommending others with relevant expertise. This strategy enabled us to capture a diversity of views representing a range of activities within and around regulation that together covered the whole spectrum. While our sample is not exhaustive, the depth and specialization of participants from diverse institutions and functions provided sufficient insight into operational practices and requirements, supporting the formulation of well-grounded recommendations for tool-makers, researchers, and regulators.

Participants were briefed on the study’s aim—“investigating the landscape of existing tools for automated dark pattern detection developed by researchers, businesses, regulators and NGOs”—and interview topics were shared in advance. Interviews lasted 45–60 minutes and were conducted between January and April 2025. Participants were not financially compensated but were offered early access to anonymized analysis outcomes.

\begin{table}
    \centering
    \begin{tabular}{r|c|l}\toprule
    \textbf{Organization type}& \textbf{Participant}  & \textbf{Expertise/}\\
    & \textbf{code}  & \textbf{Remit}\\ \midrule
 Business & BUZ-P1&law and tooling\\
 \hline
 Policy-making & PM-P2&law\\
 organization  & & \\
  \hline
Non-governmental  & NGO-P3&tooling\\
organization  & & \\
 \hline
 Consumer protection & CP-P4& law\\
  agency & CP-P8&law\\
 & CP-P9&tooling\\
 \hline
        Data protection agency & DP-P5& tooling and law\\
        & DP-P6& tooling\\
        & DP-P7& law\\
    \end{tabular}
    \caption{Participant demographics organized per organizations they work at and primary area of expertise (law includes legal enforcement, while tooling corresponds to tool development). All have direct experience with DP detection. Each participant has been identified with an acronym composed of their type of organization (BUZ= Business; PM= Policy-making organization; NGO= Non-governmental organization; CP= Consumer protection agency; DP= Data protection agency) and their number (e.g., P1), so for instance BUZ-P1 or CP-P5.
    }
    \Description{Participant demographics. This includes organization type, participant code, and the area of expertise or remit for each participant.}
    \label{tab:participants}
\end{table}

\subsection{Interview data analysis}

To analyse the anonymized interview transcripts, we used Thematic Analysis (TA) \cite{braun2021thematic,braun2021one}.
TA aligned with the need to capture the realities of the practitioners, their interactions with other stakeholders, and their needs around emerging tooling (itself considered relative to developments in digital service regulation).

We applied a codebook-style TA between two author-coders, informed by our interview questions. The effectiveness of codebook TA does not usually rely on high inter-coder agreement \cite{braun2021one} -- the primary focus here is on using the codebook to support discussions and interpretation of emerging themes. Coding was inductive, as regulatory practitioners are under-studied, and there is a need to situate understanding of automated tooling and Enf-Tech in their reported activities. We then grounded the discussion of automated tools in their experiences.

There were two coders, Coder1 and Coder2. Coder1 has a legal design and usable privacy background; Coder2 has a security usability and sociotechnical security and privacy background. Both Coders were present for the majority of interviews (except with CP-P9, interviewed only by Coder1). Both coders  made notes after each interview during anonymization, if not already after the interview itself (`Data familiarisation and writing familiarisation notes'). Both Coders coded interviews separately (`Systematic data coding') then discussed them, in batches of three, to develop and refine themes (`Generating initial themes from coded and collated data'). A coding review meeting was conducted after each batch.

Coding review meetings were more intensive during initial theme generation (`Developing and reviewing themes'). The two coders began with independent codebooks, which were iteratively merged across batches of interviews. After finalizing the codebook (Appendix \ref{app:codebook}), all interviews were re-coded. Coding meetings lasted 2–3 hours each (`Refining, defining and naming themes,'). Code groups from the final codebook informed the structure of the Results section.

Where there was disagreement in codes, this was concentrated in the first batch of three interviews, and was primarily in whether the codebook centred around practitioner activities, or the tools they used -- this already resulted in the beginnings of our process diagram (Figure \ref{fig:activities}). Resolving disagreements was then primarily an activity of deciding which code group to put a code into, more than the meaning of the code. The coders were also mindful not to `force' agreement if it was an issue of stakeholder perspective -- a few parallel codes would then be allowed to persist, to preserve multiple perspectives on the same factor.
After the second batch of three interviews, in discussion there was high agreement in what the Coders regarded as of note in the interviews -- given the rich nature of the content, disagreement was in which elements were of interest. This was resolved by developing themes and referring to the process diagram as a shared artefact. 

Coder1 saw tools including AI as the central object of study, and for Coder2 it was practitioner activities (including investigations) and utility of tools, including AI. There was then a shared view of tools and activities being strongly linked, which informed how coding and discussion of themes were developed and reconciled. 
There was high agreement in what phenomena both coders identified in the transcripts, and where they placed those phenomena in identified practitioner activities. 
Bringing together the perspectives of the two Coders was able to identify, for instance, a theme around how practitioners try some tools with new features, to discover new ways to identify signs of potential violations.

\subsection{Analysis of existing AI-based computing technologies}
Once the results were elaborated, AI-based computing technologies developed by academia and introduced in Section \ref{sec:sota} were analysed (as detailed in Section \ref{sec:reflect-tools}). The aim was not to provide an assessment of the quality of the approaches, which is discussed in due scientific venues. Instead, the goal was to understand how they are positioned with respect to the findings concerning the detection and enforcement process illustrated in Section \ref{sec:results}. In particular, tasks were identified which the different families of tools can contribute to, within the detection and enforcement process; the main functionality/ies of each kind of tool and the value of this to practitioner activities; the type of data they gather or analyse and the utility of this data for enforcement; and, whether they meet or at least consider relevant constraints and requirements of regulatory teams and their processes. We also aimed to remove the ambiguity there may be about what it is that automated tools are (believed to be) doing or that they can do, so we also wanted to have a common definition against which to discuss the views of both interviewer and interviewee.

\subsection{Ethics, privacy and confidentiality}
The study received approval by institutional Human Research Ethics Committee (HREC), including a data protection review. Interviewees gave informed consent, could skip questions, and withdraw at any time. Interviews were conducted via Microsoft Teams, recorded, transcribed, and anonymized; original recordings were deleted.
Transcripts were manually corrected for errors in e.g., acronyms, URLs, and terms. 

\subsection{Limitations}
Our participant cohort is smaller than in many studies with practitioners; participants were selected for coverage across stakeholder roles and direct involvement in detection and enforcement. Participants were highly specialized, with years of experience, and as such not trivial to reach. Findings confirm that there is limited regulatory capacity, with few experts relative to the scale of the environments they oversee.
We did not interview academic tool-makers, which we identify as future work. However, Section \ref{sec:reflect-tools} compares academic research with our findings. Our scope excluded commercial detection tools. This choice was validated by interviewees, who did not mention any such products.

The study focused on EU regulation, given the relevance of GDPR and emerging Digital Decade legislation like the Digital Services Act (DSA), and the authors' familiarity. While participants referenced other jurisdictions (e.g., California), the EU is at the forefront of dark pattern regulation. These patterns are not region-specific and pose risks globally; our findings thus inform broader discussions on fair business practices \cite{narayanan2020dark,ISOLA2025106169}. We also included participants from various EU regions to reflect differences in legal implementation and resourcing.

\section{Results: Interview analysis and tool presentation}
\label{sec:results}
This section presents our results. Each subsection represents a code group, apart from the last subsection of this section, which details broader themes which emerged from our analysis.

\subsection{Activities}\label{ssec:activities}

\begin{figure*}[t]
    \centering
    \includegraphics[width=1.0\linewidth]{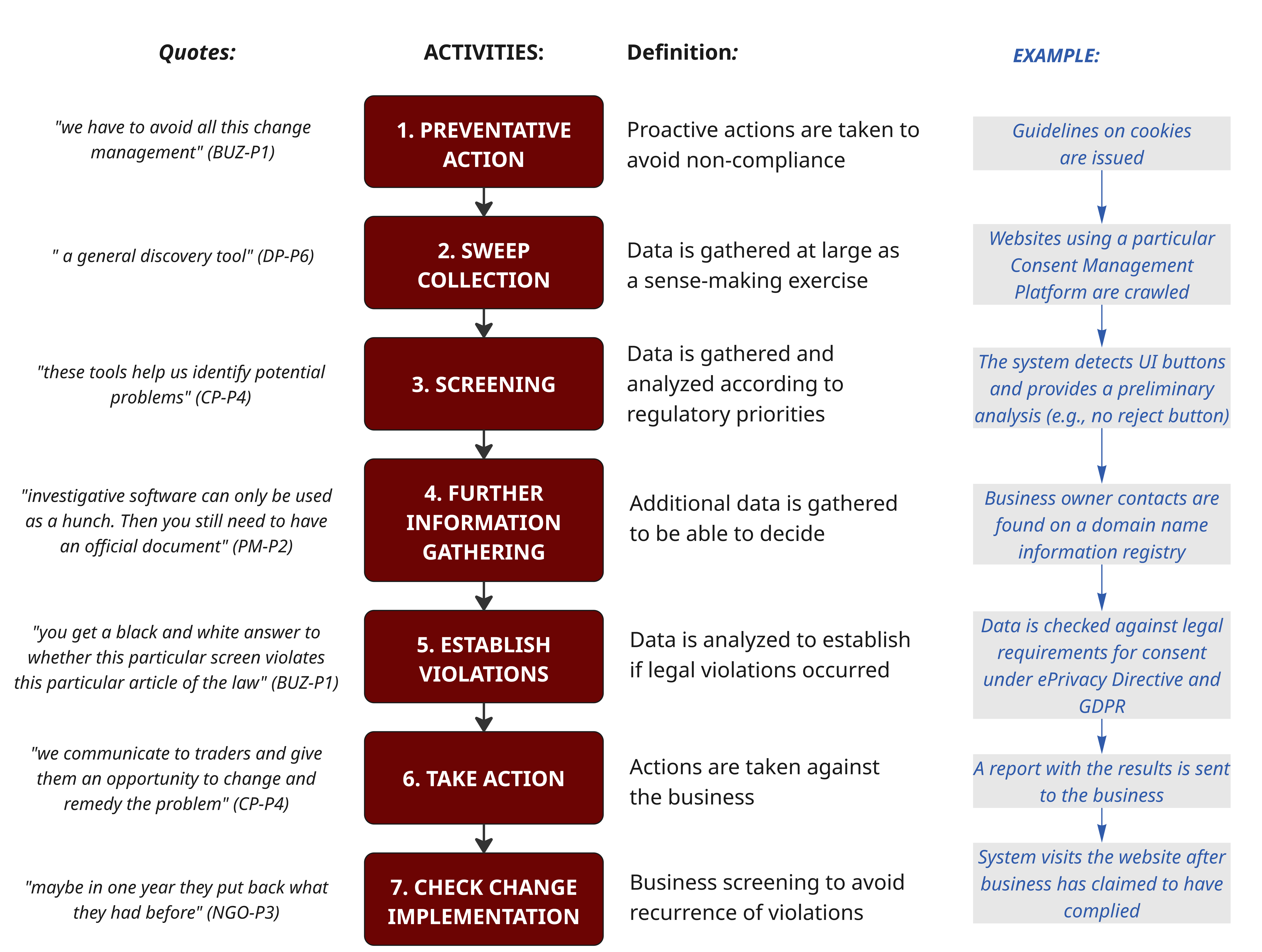}
    \caption{The phases reflecting the activities of enforcement identified in the interviews. A shorthand definition is provided for each phase on the diagram's right-hand side, and an appropriate quote on the left-hand side. On the far right is a fictional example of what the enforcement process could look like, inspired by the activities described by our participants.}
    \label{fig:activities}
    \Description{Sequence of activities identified in analysis, spanning detection and enforcement. The sequence moves through Preventative Action, Sweep Collection, Screening, Further Information Gathering, Establishing Violations, Taking Action, and Checking Change Implementation. An example is used to demonstrate the flow of activities for website cookie banners.}
\end{figure*}
Our participants span different stages of dark pattern (DP) detection and enforcement. By integrating their perspectives and tools, we constructed an overview of ecosystem activities (Figure \ref{fig:activities}), thereby answering RQ1. 
The diagram is a schematic representation of the activities mentioned, but not all stakeholders follow each step, depending on their organization's mandate. For example, all data protection authorities have the mandate to disseminate information on compliance and to issue guidance interpreting the law, but not all have done so with respect to DPs. Or, in principle, all consumer protection authorities could verify whether the requested change has been implemented, but only a few have mentioned doing so.
A key insight is the vast discrepancy between the number of online services, the DP detected, and those pursued. Regulatory action is limited to a small subset, often framed within targeted campaigns and investigations, as illustrated left-to-right in Figure \ref{fig:activities}.

Detection and enforcement involve multiple, non-mandatory phases. Practitioners use tools to automate tasks beyond detection (e.g., contacting businesses, issuing reports), while some steps require human input (e.g., legal assessment, filing complaints). Regulatory actors cannot instantly identify new services or interface changes, and must verify whether requested modifications are implemented or reverted. Some businesses may actively evade detection (e.g., by blocking traffic coming from governmental IPs, or by writing code in ways that are more difficult to parse automatically), requiring adaptive strategies. Interface scans offer limited insight: some issues (e.g., hard-coded stock levels) demand external follow-up, while others (e.g., backend manipulation) may remain hidden. Overall, the process must be traceable and auditable ("\textit{you need to capture it in order to prove that [the violation] is really happening}" CP-P4).

\subsubsection{Preventative action (Step 1)}
Before the actual enforcement activity starts, participants mentioned proactive actions that can help businesses adhere to regulations, such as issuing official guidelines interpreting the law, providing self-assessment tools or proposing consultancy services. Participants saw benefits in such pre-emptive actions, for example ``\textit{avoid[ing] change management}'' (BUZ-P1) such as redesigning interfaces and services after they have been put in use.

\subsubsection{Sweep collection (Step 2)} 
Regulatory practitioners often begin with large-scale sweeps of GUI-visible business practices, using manual or automated methods to gather broad, sometimes redundant data. These activities serve general monitoring or targeted suspicion (e.g., presence of specific design elements or platforms), aiming to assess whether illegal practices may exist ("\textit{general discovery tool}" (DP-P6); "\textit{we wanted to know if the suspect [sic: suspicion] could be proven}", CP-P9). This initial phase may involve scraping thousands of websites and can yield statistical evidence of problem scale.

\subsubsection{Screening (Step 3)} 
Even without large-scale scraping, data analysis must be guided by organizational priorities, sector focus, complaints, media attention, technological developments, or company size. Capacity constraints are central: regulators "\textit{cannot fine 30,000 businesses even in 100 years}" (CP-P4). Tools may suggest preliminary analyses and flag potential issues based on benchmarks ("\textit{if it's over 20\% then you can raise the flag}" PM-P2) or similarity to past cases.
At this stage, accuracy is secondary; tools provide indicators of possible violations "\textit{tools help us identify potential problems}" (CP-P4), which require deeper investigation. As one practitioner noted, "\textit{we have a very good indicator for where trouble may be occurring, and then... you would still have to dive in deeper to actually determine what's happening, and to determine whether something is illegal or not}" (CP-P4).

\subsubsection{Further information gathering (Step 4)} 
To establish legal violations, regulators often require data beyond what is collected online. For instance, verifying whether a low stock message reflects actual inventory cannot be done by the tool alone.  Additional information may be obtained from businesses or through resource-intensive audits, as detailed in Section \ref{ssec:data}. This data supports subsequent steps, such as identifying service owners: "\textit{investigative software can only be used as a hunch. Then you still need to [...] contact the [website] registry and you need to get an official certified paper}" (PM-P2). This is challenging since the identity and contact details of business owners "\textit{is most of the time very well hidden}" (NGO-P3).

\subsubsection{Establish violations (Step 5)} 
Matching potential dark patterns to legal violations is complex due to legal subtleties and always requires human validation. Automated tools assist but cannot replace expert judgment: "\textit{so many factors are at play that [...] it was not efficient to build a system that would be able to do that for us}" (CP-P4). To establish legal infringements, there is the necessity to "\textit{get a black and white answer to whether this particular screen violates this particular article of the law}" (BUZ-P1). For the NGO, human involvement is also essential to file complaints, which must concern individuals and cannot be filed directly by them.

\subsubsection{Take action (Step 6)} 
Once violations are confirmed, regulators may pursue complaints or fines, though smaller businesses are often given a chance to redress issues first: "\textit{we communicate to traders and give them an opportunity to change and remedy the problem}" (CP-P4). Some participants "\textit{address letters to companies, ask them to comply and [tell] them in which way they can comply}" with fines issued only if non-compliance persists (DP-P7). Severe cases may bypass warnings. NGOs and businesses may suggest countermeasures, but some authorities avoid this to prevent endorsing potentially non-compliant solutions. Letter content can be tool-generated, especially for smaller businesses who may unknowingly implement DPs made available by third-party providers.

\subsubsection{Check change implementation (Step 7)} 
After enforcement, some regulators and NGOs continue monitoring services to verify implementation of requested changes and detect potential reappearance of dark patterns over time: "\textit{maybe in one year they put back what they had before}" (NGO-P3).

\subsection{Tools}\label{ssec:tools}
Participants used a range of software tools, especially in early enforcement steps (e.g., data collection and narrowing targets). Later stages shifted toward legal analysis, where tool capabilities diminish and data quality becomes central.

\textit{Tools for evidence collection} included scrapers, artificial agents, sock puppets, virtual machines, and VPNs for anonymous browsing. Violations were documented via screenshots and video recordings of user journeys. Interface scanners detected UI elements like cookie banner buttons and contrast ratios.

\textit{Tools for data analysis} covered various AI and data science methods, such as techniques for statistical analysis and semantic similarity algorithms comparing contractual clauses with case law. These tools flagged potential issues (Step 3), but legal violations were assessed manually (Step 5), since tools cannot do that kind of fine analysis. 

Since fines require a legal ruling, any data moving to later analysis stages must be credible enough to withstand judicial scrutiny. Otherwise, it serves only as guidance. Consequently, participants stressed the importance of anticipating what judges might need, in order to simplify later steps. This is why, for instance, we collected mixed opinions about the use of LLMs: while two participants praised their reasoning capabilities, others were more sceptical.

\subsection{Data, evidence and dark patterns}\label{ssec:data}

\subsubsection{Data} Participants described a range of data types collected for professional purposes, some of which serve as evidence of potential legal violations, often linked to DPs. Typically, they gather publicly available data (published on websites, portals, or by authorities as open data) captured via time-stamped screenshots or recordings, sometimes continuously (e.g., countdown timers), as noted by BUZ-P1, CP-P4, and PM-P2. 
Confidential data may also be collected during investigations, reflecting the diversity of evidence types. This includes information from `dawn raids' (unannounced inspections of a company's premises), verification visits, written requests, interrogations, complaints, and expert input.

Importantly, in-depth investigations require linking interface data to other sources of information, sometimes through compelled data-sharing or surprise inspections. The process must align with legal definitions: “\textit{you don't care whether it's called a dark pattern or not}” (BUZ-P1). Thus, most participants focus on legal violations rather than interface patterns. As DP-P6 explained, “\textit{as a legal officer, [...] what gives us leverage [...] is to have something that matches with the legal definitions [...] Then [dark patterns] is not a term that we actually use, but it's a really useful term for studies.}” Similarly, CP-P4 stated, “\textit{even the mentioning of a dark pattern doesn't mean anything to me. In the end, what I'm interested in is something that violates the law}.”

\subsubsection{Dark patterns}
Participants referred  both to dark patterns and legal violations. Explicitly mentioned patterns included countdown timers, misleading discounts, price discrimination, fake reviews, urgency, social proof, hidden costs, confirm-shaming, and low-stock messages. They also referred to GUI elements (e.g., buttons), technical components (e.g., Identifiers for Advertising, HTML elements), and user journeys such as subscriptions, cancellations, and booking flows. Others emphasized legal issues like transparency, uninformed consent in cookies, and abusive contractual clauses.

Definitions of DPs at a higher level varied. BUZ-P1 linked them to regulatory needs, describing them as interfaces that manipulate users against their interests. DP-P6 argued that intention to mislead must be “\textit{striking}.” CP-P8 mentioned using legal knowledge, policy reports, neural marketing tests which measure bodily response of participants exposed to DPs (such as stress), and consumer surveys to determine whether an interface is manipulative. Cultural factors also emerged, such as users being “\textit{trained to accept}” (BUZ-P1) and experiencing “\textit{helplessness}” (DP-P6). Conversely, NGO-P3 observed that some service owners do not treat GDPR violations as legal concerns.

Some participants clarified also what they do \textit{not} consider DPs, even though this distinction is not always clear-cut. As DP-P6 noted, legal requirements are “\textit{very clear for us},” but interfaces are subtler, especially when obligations are vaguely defined. DP-P7 added that “\textit{it's quite subjective [...] we don't take the risk [...] if we don't have strong evidence}.” Uncertainty also stems from limited case law: “\textit{case law is never as clear cut as we want it to be}” (CP-P4). CP-P8 described relying on prior judgments to match new evidence, while DP-P6 saw future potential in automating detection using “\textit{objective markers}” of the presence of dark patterns.

\subsubsection{Matching dark patterns and their regulation}
Linking manipulation directly to legal violations is complex and often depends on prior case law. Enforcement may involve pattern-matching, but only for clear cases like missing ‘reject’ buttons or textual expressions that are similar enough to a given dataset. Regulatory actions are primarily grounded in established data protection laws (the General Data Protection Regulation and the ePrivacy Directive) and consumer protection (Unfair Commercial Practices Directive and the Price Indication Directive, which was notably highlighted despite its limited presence in DP literature), and newer instruments of the EU Digital Strategy (like the AI Act, the Digital Services Act, the Digital Markets Act). References to the California Privacy Rights Act and the US Federal Trade Commission also emerged, although our focus was European.

\subsection{Requirements and constraints for enforcement tooling}\label{ssec:constraints}

To answer RQ2, participants identified several constraints and requirements for using automated tools in regulatory practice, particularly regarding transparency, traceability, and result accuracy, which are essential when data may enter legal proceedings. Moreover, automation was seen as a way to reduce investigation costs during early detection phases, ahead of enforcement. Related to these dimensions, data confidentiality and open-source software were also discussed as important requirements.

\subsubsection{Trustworthiness for accountability} \label{ssec:trust}
As previously noted, regulators cannot feasibly scan all digital services and must select a subset. In some jurisdictions, this selection must be \textit{transparent} so as not to be seen to be unduly targeting a specific business. CP-P4 warned that some practices may be targeted simply because “\textit{some tools may be easier to build than others},” risking misalignment with legal standards of impartiality.  Moreover, ensuring the integrity of analysis results is crucial for establishing a “\textit{chain of proof}” (DP-P6). Participants emphasized the need for accountability in tool development, whether through trusted institutions, international adoption (e.g., EDPB Support Pool of Experts\footnote{\url{https://www.edpb.europa.eu/support-pool-experts-spe-programme_en}}), or open-source access to code (see below).

The \textit{reliability} of automated findings was a recurring concern, but tools must operate with high certainty to be trusted. Yet, as discussed in Section \ref{ssec:data}, it remains difficult to consolidate objective knowledge about DPs which can be automated. Tools often capture only surface-level features and struggle with “\textit{complex analysis}” (DP-P6). Without contextual understanding, they risk producing inaccurate results, especially as DPs are continuously developing.

To ensure a “\textit{bulletproof process}” (DP-P6) that businesses cannot challenge, \textit{verifiability} was regarded as crucial. In legal proceedings or appeals, all data must be accessible for external review, and the procedure must be “\textit{correct},” “\textit{non-biased}” (CP-P4), and based on “\textit{sound logic}” (DP-P6). For example, browser choice was highlighted as key to reproducibility in cookie scanning. CP-P9 also cited legal obligations for public administrations to release “\textit{explainable AI}.”

Concerns were strongest regarding opaque technologies, such as generative AI, due to uncertainty about their admissibility in court and variations in evidence rules across Member States. Trust in computational tools remains fragile: “\textit{it's very hard for us to take machine learning models and LLM output at face value}” (DP-P6). While LLMs offer flexibility, hallucinations were reported, e.g., a tool wrongly flagging missing information. Conversely, CP-P8 found it “\textit{quite surprising}” that an LLM “\textit{found more [possible] dark patterns than humans}” on a set of websites where a new tool was being tested, comparing its findings with the DPs that the staff had identified. CP-P8 was optimistic on the tool's potential to strengthen their detection abilities, while avoiding over-reliance: "\textit{Sometimes we didn't agree with this assessment, but in many cases it was really the good direction to have some more issues to consider}." \textit{Traceability} was also considered a necessary requirement, especially when tool inputs can be modified or when human intervention is needed to resolve inconsistencies.

To enhance trustworthiness, participants mentioned documentation (DP-P6), removal of incorrect results (CP-P4), and staff training (CP-P8). However, caution was advised: “\textit{a certain degree of modesty}” (CP-P4) is needed to acknowledge tool limitations and “\textit{consider very carefully [...] what we can do with such a tool}” (CP-P8).

Lastly, human oversight was considered essential to ensure accountability, as “\textit{it's easier and safer to do that by humans}” (CP-P4). No participant believed that legal processes could rely solely on automated tools; instead, tools were seen as supportive, with decisions remaining firmly in human hands. As DP-P6 put it, “\textit{The one who makes the decision is always the human, so we don't depend on the tool 100\%},” especially since tasks like establishing violations cannot be automated.

\subsubsection{Cost}\label{ssec:cost}
We observed a general trend toward work optimization enabled by technologies such as assisted writing and semantic search. These tools helped “\textit{do much more with a lot less capacity}” (CP-P4), saving time across tasks such as screenshotting, timestamping, organizing data, and retrieving internal case law. However, upfront investment was often required, making off-the-shelf solutions attractive (see Section \ref{ssec:dev}). At the same time, concerns emerged about data overload, especially when authorities lack capacity to process it. As PM-P2 noted, “\textit{you still need to analyse them one by one}” to ensure reliability and accountability before legal proceedings.

\subsubsection{Open Source}\label{ssec:open-source}
Open source software and APIs were considered essential by DP-P5, DP-P6, and CP-P9 for transparency and reproducibility. Open source also enabled sharing tools with other authorities, as it allowed modifications and “\textit{full control of how it works}” (DP-P5). However, maintenance costs were seen as a barrier to release.  Open source further implies being free, which cuts the decisional and organizational burden of public procurement: "\textit{in a public administration [...] if you have to buy software you have to go through very difficult procedures}" (CP-P9).

\subsubsection{Confidentiality}\label{ssec:confid}
Participants emphasized the need to protect investigation-related data from tampering or leaks and to store it securely. To prevent unintended use, CP-P8 recommended closed environments for LLMs, and CP-P9 avoiding training or fine-tuning with sensitive input. Confidentiality was also seen as a barrier to collaboration with external partners, which could otherwise help distribute the cost and effort of developing enforcement tools.


\subsection{Co-development and collaboration with academia}\label{ssec:dev}
BUZ-P1 distinguished between regulators with in-house technical teams and those relying on external support due to limited capacity. The latter often need to first find out what is feasible “\textit{from the technical point of view}” (CP-P8). Regardless of setup, collaboration with legal experts is essential, typically involving small “\textit{feedback loop[s]}” (NGO-P3) and resource investment, such as building training datasets. Academic partnerships were seen as valuable, especially through spin-offs that commercially leverage intellectual property created within universities. Research was praised for its insights into IT possibilities and expert approaches, though some felt it lacked a “\textit{wider picture}” (CP-P4), despite being “\textit{always incredibly useful}” (DP-P6).

\subsection{Critical points for future developments}

Just as constraints were identified regarding what automated enforcement tools can achieve within regulatory activities, specific areas of potential were also explored, particularly during the latter part of the interview, to determine if certain barriers could be addressed.

The sheer abundance of websites, and for example, cookie banners, was seen as providing plenty of training data for large-language models.
However, the potential of automation is foremost in screening rather than in detecting specific legal violations: "\textit{without objective indicators [...] that would fit to our doctrine [...], it would be more of a general discovery tool and} \textit{not very applicable in terms of automated enforcement}'' (DP-P6). It may be feasible to compare interfaces to prior judgments to indicate potential violations, as happens with unlawful contractual clauses, but this requires there to be a sufficient amount of, and sufficient clarity in, prior cases.  

Another type of undue influence on consumers that is worth investigation is "\textit{when people's concern for the environment is exploited}" (PM-P2). There is also certainly room for development in web tracking, particularly in the identification of unique identifiers. However, concerns were raised about the actual capabilities to do so when identifiers are created on the server side of the business, which authorities will need to access directly, rather than the user browser that can be more easily inspected. In general, the capability to scan visible website content was seen as improving, but that if businesses wished to avoid legal scrutiny, they may move their deceptions to the back-end.  

Hope was expressed for off-the-shelf data science methods that can be used to analyse confidential data.
With the advent of trained autonomous operations browsing for users, it was considered that automated agents could be trained not only to detect manipulations and adapt in order to avoid ``\textit{falling into these traps}'' (BUZ-P1), but also to uphold a user's configured preferences. 

Some recommendations were explicitly given by the participants, mainly focusing on automating screening activities or market monitoring, for example to build applications that scan massive numbers of websites and compare them to quantifiable indicators of violations. Importantly, CP-P9 advises contacting the data science unit of the relevant authorities to co-develop tools for specific cases or sectoral inquiries that the agency is pursuing. Tailoring is necessary because "\textit{those tools must fit the actual work of the authority}" (CP-P9).

\subsection{Broader themes}

Part of the thematic analysis process \cite{braun2021one} is to identify broader themes, primarily by (reflexively) leveraging the authors' backgrounds to identify critical findings across code groups. These broader themes are summarized here.

\subsubsection{Detection tools as inspiration.}  Easy-to-access tools may be used to explore what is possible and inform the development or deployment of in-house tools (Section \ref{ssec:tools}), whose data and results must be protected from outside view. In this sense, available tools inspire formal investigation processes, but should not be expected to be used directly as part of legal procedures, at least not for the moment (relating to transparency, Section \ref{ssec:trust}). This limits their use, but to something useful. Detection tools have the potential to augment regulators' experience, given that they are not necessarily UX designers (with some capacity to imagine how different interface designs may manipulate users). This is in essence how new tools draw their attention, providing a way to detect deceptive patterns which are new to them. The law is abstract in terms of defining design rules, and in that sense can be violated in many ways, but interfaces have variations too. We found that a practitioner's imagination on one side moves in step with their experience of the other, regardless of their role relative to regulation.

\subsubsection{Campaigns rather than coverage.} Investigations are focused on specific 'families' of dark pattern or even individual ones, typically because they match clearly with specific legal violations. Regulatory activity is usually more focused on campaigns than coverage. It involves having substantial evidence of specific infringements rather than weak indicators of a range of violations (as discussed around tooling, Section \ref{ssec:tools}), where there is not necessarily a similarity across cases. CP-P4 was cautious about the possibility of reaching broad coverage automatically since "\textit{[the tools] would have to be so specific, but also the deviations between different cases are so subtle, that I wouldn't really know if it's possible to program it}".
The detection range of the tools that aim for coverage of a broad set of DPs (such as \cite{shi202550,Deceptilens_2025, mansur2023aidui, chen2023unveiling}) was considered too general to be of actual use, because agencies work on specific cases or sectors, driven by various motivations, reported in Section \ref{ssec:activities}. When there is no match in the scope of the analysis capacity of existing tools, it is difficult, if not impossible, to find a "\textit{clear usage}" (DP-P6) for them. 

\subsubsection{Scaling the diligence, as much as the detection.} AI-supported tools are foremost seen as a way to scale up and match the prevalence of dark patterns online, rather than to augment the skills of regulators (as seen across Sections \ref{ssec:activities} and \ref{ssec:tools}). The process needs to be transparent and legally sound from the start (Section \ref{ssec:constraints}). 

Tools that make data management easier are appreciated for the unique features they possess, namely the ability to detect particular kinds of dark patterns or infringements and catalogue them efficiently. Where automated-detection tools hold promise is in the time-intensive and repetitive tasks of visiting large quantities of websites and logging details of each interface, although most illegal practices cannot be detected by interface interrogation alone: often the code needs to be inspected (e.g., to understand whether a countdown timer resets automatically after a preset time) and additional information needs to be gathered directly from the business. This constitutes evidence on its data or consumer practices that is sufficient and appropriate to uphold a legal claim. 
Automated tooling then has potential foremost in indicating clear infringements of regulations at scale, provided that existing case law is unequivocally matched to DPs and that reliable datasets are available. That is to say, the comparatively small regulatory teams cannot scour the whole internet, but automation may help to catch up with `obvious' infringements that were existing in plain sight up until that point. This is, however, dependent on the pace at which automated tools can keep up with efforts to evade detection. 
Moreover, an essential constraint is the internal capacity of enforcement agencies: if technologies provide an overload of data without filtering or organizing it, this may hamper rather than facilitate their work. Similarly, some participants highlighted the efforts required to build databases and train computing tools, especially as an initial investment.

\section{Results: Reflection on Computing Technologies}
\label{sec:reflect-tools}

In this section, we present the results of our analysis of AI-based computing technologies proposed by academia to answer RQ3, in particular to determine: to what extent the phases of enforcement activities are covered (Section \ref{ssec:activities}); whether the tools serve for evidence collection or for data analysis (Section \ref{ssec:tools}); the type of data that is gathered and/or analysed, with a focus on whether it is matched to legal requirements or violations (Section \ref{ssec:data}); and, whether they meet or at least consider relevant constraints and requirements (Section \ref{ssec:constraints}), such as considerations about the trustworthiness of the approach, including if the data and the code are available (which is paramount to verifiability and transparency).

First, we notice that all the analysed tools \cite{mathur2019dark, Nouwens2020, kirkman2023darkdialogs, chen2023unveiling, mansur2023aidui, bouhoula2023automated, gundelach2023cookiescanner, Deceptilens_2025, shi202550} are useful for the Sweep Collection phase (as in Figure \ref{fig:activities}), where there is not a particular regulatory priority that is set, but it is mostly an exploratory activity. 
Many of these approaches \cite{mathur2019dark, Nouwens2020, kirkman2023darkdialogs, bouhoula2023automated, gundelach2023cookiescanner, shi202550} include web crawlers that gather data from websites, online portals and mobile apps, mostly GUI data, but also other types of data such as cookie content. These tools would be useful for evidence collection, but the data they collect is usually not timestamped. However, logging details of data gathering is a crucial feature for regulators; this is then an example of where evidence collection must consider the fuller enforcement process, with data as potential evidence. Other approaches \cite{chen2023unveiling, mansur2023aidui, Deceptilens_2025} derived their datasets from existing datasets of generic graphical user interfaces containing a subset of DPs. Such data, however, is not representative of legal infringements.

Only a few approaches \cite{Nouwens2020, kirkman2023darkdialogs, bouhoula2023automated, gundelach2023cookiescanner} can benefit regulators in those phases that are more focused on  regulatory objectives, such as Screening and Establishing Violations. Indeed, these approaches all focus on automatically finding violations on cookie banners, exemplified by various types of DPs. Even if not explicitly related to legal violations, some other approaches \cite{mansur2023aidui, Deceptilens_2025,chen2023unveiling} may nevertheless be useful for data analysis, if the focus is on DPs. Then, there needs to be a phase in which these are mapped to legal infringements, to justify the subsequent use of resources to examine them once they are added to a caseload. 

Most of the analysed approaches show attention to the scientific reproducibility of the studies, uploading their source code and/or the datasets to online repositories such as Zenodo and GitHub. However, only a few declare the use of free and open source software with a copyleft license \cite{mathur2019dark, kirkman2023darkdialogs, bouhoula2023automated}. This shows attention to their verifiability, one element that was vital for our participants. For many others, the license was not specified. The LLM-based approaches \cite{Deceptilens_2025, shi202550}
used proprietary models as a baseline because of their superior performance to open source models, which, even if they improve the accuracy of results, do not make them attractive to regulators, because they do not provide the necessary safeguards in terms of confidentiality of the data, and the verifiability and traceability of results. 

Only very few tools explicitly addressed other dimensions related to trustworthiness of the system, with \cite[p.17]{chen2023unveiling} underscoring the importance of "monitoring and maintenance [...] for responsible AI practices", and \cite{bouhoula2023automated} for human oversight. \cite{Deceptilens_2025} explicitly addressed opaqueness and accuracy problems, by implementing RAG (Retrieval Augmented Generation) and involving DP experts in the evaluation of the LLM's output for transparency reasons.

When it comes to considerations about the trustworthiness of the approach, we notice that only a few works devote attention to it. For example \citep[p. 1733]{bouhoula2023automated} recognizes the risks of inaccurate results and the need for human oversight: "the violations observed by our automated procedure cannot be directly taken by a court or DPA to enforce fines. Given the risks of false positives, one must inspect and confirm them manually", also due to "ambiguous interpretation" of data (composed of natural language, cookie content and cookie interface elements).

Regulatory practitioners may use additional digital tools in a speculative manner (e.g., VPNs, code-viewers in a browser, etc.). There is scope to move some of this work into the sweeps, but that would only move effort from the step of Further Information Gathering to the earlier steps of Sweep Collection and Screening; the effort is automated, but there would still be later human effort to verify what was gathered, rather than perform the gathering, thereby transforming one form of effort into another. %

Referring back to Figure \ref{fig:activities}, most research tools focus on the step of Screening, and with varying strength. The Filtering and Prioritizing is reliant on human effort. This represents a divergence in the grander aims of both communities. It may look good for tools to run large-scale sweeps and be seen to collect large amounts of evidence. 
However, a connection is required between a bad practice and a legal requirement, for a detected dark pattern to reach the \textit{end} of the process. Dark pattern detection tools then work better as warnings to take action. 

Another bottleneck is that tool developments would rely on case law moving at the same pace as detection capabilities. 
Automated tools that produce weak signals may still be useful for getting a first sense of an emerging bad practice. 
This still means though, that thorough investigation has to follow before action can be taken. 
All the while, not all bad practices can be detected by only looking at an interface. There is then a wall that automated tooling will hit and cannot get around.

\section{Discussion}
\label{sec:disc}
Based on our findings, we provide initial recommendations for next steps to align the pace of automated enforcement technology to the needs of intended users.

\subsection{Map jurisdiction-specific case law to DPs and co-design technologies that identify potential legal violations upfront}

Automated tools are of growing interest to regulators, businesses, and NGOs, not only in consumer protection enforcement \cite{Riefa_Coll_2024}, but also in data protection and competition law. A key challenge, however, lies in scaling the detection of deceptive digital practices in line with legal definitions: whereas most existing approaches focus on interface-level analyses, translating these into legally grounded claims requires additional work that the community has only sparsely engaged with. Only a few tools explicitly map dark patterns to legal provisions \cite{Nouwens2020, kirkman2023darkdialogs, bouhoula2023automated, gundelach2023cookiescanner}, whereas all our interviewees need to connect weak signals to plausible legal violations, and those to existing case law.

Automated tools need some capability to map possible dark patterns directly to legal requirements, prior legal cases, or blacklists of inappropriate practices.
However, this mapping is particularly difficult for certain DPs, because criteria to objectively define whether an interface contains DPs are lacking. In consumer protection, pattern-matching benefits from blacklists of clearly defined illegal practices (e.g., in the Unfair Commercial Practices Directive) and established case law. Indeed, CP-P8 describes a semantic similarity tool for spotting unfair contractual terms based on past jurisprudence. This clarity, though, is absent in other domains, especially under very recent laws such as the Digital Services Act, the Digital Market Acts and the AI Act in the EU, where the absence of case law causes open-ended interpretations of what may be admissible. This mirrors issues seen in, e.g., security compliance expectations in organizations \cite{woods2021blessed}, where what is reasonable is made clearer through (slow) case law. For DPs, there are instances where the mapping is not clear-cut due to scarce case law, but existing cases can be leveraged to reduce the reliance on regulatory practitioners to do that mapping.  

In general, many academic tool-makers are not in direct engagement with legal frameworks, often due to exclusive familiarity with DP academic terminology or the international scope of their work that transcends jurisdictions. While legal provisions are abstract and broadly applicable, DPs are highly contextual and varied in how they can infringe the law. Pattern-matching approaches typically focus on specific interface designs instead of dynamic interaction flows that, however, may better align with the narrative structure of case facts  \cite{Gunawan_2025} and are limited by that scope, thereby lowering their scalability. On the contrary, enforcement agencies rely on implicit mappings between design elements and legal norms (e.g., the presence of countdown timers signals a potential legal infringement), which are not necessarily formalized during the initial Sweeping activities, but only in later stages of the investigation.

Interdisciplinary and inter-sectoral efforts are needed to map known legal violations from case law databases to known dark patterns, building on work by \cite{leiser2023dark, brignull-2023}. Even though this can be a starting point, we need to align the pace of DP adoption, especially in emerging technologies, with the progression of case law on DPs.  For instance, there is no browsable, reusable, interoperable library of administrative fines where data protection violations\footnote{E.g., \url{https://www.enforcementtracker.com, https://gdprhub.eu/} etc.} are linked to typologies of DPs. This mapping must be jurisdiction-specific, though striving for international frameworks is desirable. Outputs should be computable and interoperable (e.g., via formal legal ontologies \cite{pandit2019gconsent, palmirani2018pronto}) to support the automation of compliance checking. Automation in legally-sensitive contexts would also require the employment of appropriate technologies, such as LLMs fine-tuned for legal reasoning, rather than general-purpose ones. The DP community has already successfully combined HCI and law \cite{gray2024mobilizing,Bielova_Santos_Gray_2024,Gray_2021interaction,Santos_Rossi_2023}, bringing the phenomenon of DP under the spotlight of research and at the forefront of policy-making, legislation and enforcement. Now, it needs to involve computer scientists more effectively, especially experts in legal knowledge representation and computational models of legal reasoning. 

More in general, tool developers should collaborate with legal scholars or practitioners early on in the development process, to align the technical requirements for data collection and analysis with legal provisions. Our interviews show that regulators' primary goal is justifying with a plausible narrative why a design pattern is a possible legal violation, rather than only finding DPs. This echoes the limitations of leveraging results from HCI studies identifying DPs and their effects on users, since regulators nevertheless need to assess the evidence as a matter of procedure \cite{Gunawan_2025}.
Artefacts such as Gray et al.’s taxonomy may serve as a boundary object \cite{gray2024ontology}, facilitating dialogue between HCI and regulatory communities. Establishing interim terminology and exchangeable artefacts can create a `trading zone' \cite{galison2010trading}, as seen  for other areas of security and privacy involving multiple communities of practice \cite{spring2017practicing}. 
While calls to align legal guidance with empirical DP research  are not new \cite{Bielova_Santos_Gray_2024}, this study identifies regulatory activities where misalignment is most obstructive. For example, whereas automated tools may substantially support practitioners' work in data collection tasks, they often lack the necessary logging of information such as timestamps, for actual legal evidence collection.

\subsection{Identify enforcement actors and their specific needs, and engage with them}
It is paramount to identify specific users and their specific needs, then actively engage with them. 
For example, regulators often use automated tools for rather simple, targeted tasks (e.g., detecting specific GUI elements), guided by organizational priorities or media attention. In contrast, academic tools aim for broader DP detection, driven by principles of generalizability.
Researchers and developers must identify relevant stakeholders and model their diverse needs into user requirements, by scrutinizing the public documents where organizational principles (e.g., proportionality, transparency, impartiality, etc.) are set and by involving prospective users in interview studies, like we did. In our experience, enforcement actors are willing to participate for the common good.

Table \ref{tab:requirements} lists important factors to consider when designing tools for enforcement, based on our results. Tool functionalities must align with operational requirements, such as institutional cost capacity, particularly the constraints faced by regulatory bodies, as they operate under significant time pressure to keep up with the fast pace of emerging digital threats, but meet the slow complexities of rigid administrative procedures such as procurement and deployment decisions. These limitations may restrict their freedom of choice regarding technological tools. Where we have identified bottlenecks in specific enforcement activities, these can become actionable information to direct tech development efforts for other stakeholders which have downstream requirements that need to be anticipated in advance (such as bolstering evidence and having a transparent data-collection trail). This may apply to, e.g., consumer advocacy groups, trade associations, retailer groups, and communities dedicated to anti-scam activities, among others.

To be of real utility, automated tools for DP detection need to include transparency mechanisms for ensuring human oversight. Open source, as mentioned below in Section \ref{ssec:pathwa}, can contribute in this sense, besides cutting costs and time of procurement processes. Adequately protecting data (see Section \ref{ssec:balance} below) that is gathered, used for model training or shared across collaborating organizations is another essential requirements.

\begin{table}
    \centering
    \begin{tabular}{r|l|c}
    \hline
     \textbf{Considerations} & \textbf{Examples from our results}  & \textbf{Section} \\
     \hline
      Enforcement   & Preventative action & Sec. \ref{ssec:activities} \\
      activity/ies & Sweep collection & \\
      & Screening & \\
      & Further information gathering & \\
      & Establish violations & \\
      & Take action & \\
      & Check change implementation & \\
       \hline
     Tool functionality  & Data analysis & Sec. \ref{ssec:tools} \\
       & Data collection & \\
       \hline
    Trustworthiness  & Transparency & Sec. \ref{ssec:trust} \\
       & Reliability & \\
         & Verifiability & \\
         & Traceability & \\
         & Human oversight & \\
         \hline
      Cost     & Time & Sec. \ref{ssec:cost} \\
       & Financial resources & \\
       & Human resources & \\
         \hline
     Open source & Gratuitousness & Sec. \ref{ssec:open-source} \\
     & Transparency & \\
     \hline
     Confidentiality & Data collection & Sec. \ref{ssec:confid} \\
     & Training data & \\
     & Collaboration & \\
     \hline
    \end{tabular}
    \caption{The first column of this table summarizes the elements that our interviewees considered when adopting computing technologies for enforcement. The second column lists core properties and examples that emerged in our data, but that can be adapted depending on the use case. For instance, \textit{cost} is an important consideration, which can be declined in terms of time-saving, expenditure, or human capital that is needed for development or use of the tools. On the right-hand side is the section where each topic has been presented.}
    \Description{Table showing considerations for the needs of regulatory practitioners when adopting computing technologies.}
    \label{tab:requirements}
\end{table}

Tool makers should also consider the heterogeneity of institutional capacities: some agencies have internal capacity to develop or adapt tools and ML models, while others rely on legal experts with limited technical literacy. This is especially relevant for AI-based applications, as in the EU the AI Act's Article 4 now mandates AI literacy for employees, tailored to their "technical knowledge, experience, education and training and the context" of use. This requirement creates a significant opportunity for HCI researchers, user-centred designers, and digital educators to devise best practices and experiment with effective training strategies that address the specific needs of this diverse population, such as using AI tools safely in legal proceedings, making an informed choice on their fitness-for-purpose and interpreting the results.

Ongoing development and  deployment of accessible dark pattern detection tools remain essential. These tools often serve as “scratchpads” or sources of inspiration for actors of the regulatory space, helping them explore relevant technologies. Tool-makers must decide early how to implement transparency requirements in the design of their tools, to promote the full accountability of decision-making in public administrations and make them enforcement-ready, or whether they will serve primarily as exploratory aids in the Sweep collection or the Screening phases. 
Currently, unlike the approaches focused on cookie notice compliance \cite{nouwens2022consent,kirkman2023darkdialogs,bouhoula2023automated,gundelach2023cookiescanner}, that are limited to this use case though, most tools \cite{mansur2023aidui, Deceptilens_2025,chen2023unveiling,mathur2019dark,shi202550} fall into the latter category: useful for inspiration but not integrated into formal processes unless adapted for regulatory engagement. Research tools often claim utility for regulators, but their actual use is constrained, because they do not align with their evidentiary needs and investigatory methods - a limitation also recently noted in HCI research on DPs \cite{Gunawan_2025}. Although the surveyed approaches are well-intentioned, a rigorous analysis of user requirements is indispensable.  

\subsection{Design for enforcement activities beyond the mere DP detection}

Our findings show that there are a number of enforcement tasks where academia can meaningfully contribute beyond strict DP detection, and this opportunity should be capitalized.
Participants emphasized the need for diverse tools, such as web crawlers and systems that record and timestamp user interaction data. Monitoring must span time, as violations may not be evident from a single screenshot, thereby casting doubt on the usefulness of tools that can only take graphical UIs as input. Since collecting more data does not guarantee legal action due to the limited capacity of regulatory agencies, academia could additionally focus on organizing, labelling, prioritizing and exploring evidence, including helping to optimize the management of complaints as also proposed by \cite{Sartor2025}, thereby freeing up resources for NGOs and regulatory agencies to run investigations.  
Moreover, research not explicitly framed as DP-related but which may be considered as such, e.g., automated detection work on complex legal language and misleading statements \cite{lippi2019claudette, rahat2022your}, can also support enforcement activities. 

Further, while scanning visible content is improving, businesses may shift deceptive practices to the back-end to avoid scrutiny. Many research studies demonstrate, for instance, that invisible data flows occur at scale regardless of users' choices on the interface and of the declared data practices, not only on websites \cite{Liu_2024, matte2020cookie}, but also on mobile \cite{Paci_2023} and IoT devices \cite{Edu_2023}. This aligns with Leiser \& Santos’ concept of “Darkest Dark Patterns” \cite{leiser2023dark}, who noted that enforcement agencies struggle more with back-end inspection than interface-level analysis, which may result in an under-enforcement issue. However, going beyond the visible interface requires specific technical expertise, such as performing dynamic analysis of apps to uncover their interactions with their environment \cite{Nutalapati_2020}. Furthermore, the evolution of technologies and business models, such as the phasing out of third-party cookies and the increasing reliance on unique server-side identifiers, is also hampering transparency and the ability of agencies and NGOs to inspect them. Speaking of this,  DP-P6 expressed concerns on their "\textit{loss of remote enforcement capabilities}'' and the independence of the agency "\textit{in the ability to check first-hand what is happening}", thus the need for more privacy-enhancing technologies (PETs) to protect the user as a result of this shift.

Future work should bridge the misalignment between what is available in academia and what is needed in enforcement. A first step includes a structured survey of existing technologies, including an assessment of their maturity and the identification of critical gaps. For instance, despite significant advances in usable privacy research, many solutions remain theoretical, with limited impact on regulatory practice.  It is paramount to additionally consider the organizational workflows where tools will be used, and the applicable laws and policies.  We recommend that the DP community confronts these key questions: how might we design processes to select, develop, and validate tools that directly support enforcement goals and respond to user needs? How might we assess them with combined HCI, computer science and legal metrics\footnote{E.g., with \textit{legal design} metrics \url{https://justiceinnovation.law.stanford.edu/projects/ai-access-to-justice/metrics/}} so that they can be safely adopted within administrative procedures and judicial systems, starting from the requirements we elicited in this work? How might we make them easily findable, browsable and reusable, and ensure their uptake by the intended users? How might we "collaboratively design technologies, social  practices and policies" in reality, as asked in recent CHI research to bridge HCI and policy \cite{Yang_2024hci-policy}? 

\subsection{Establish viable tech adoption pathways and shared spaces for co-design of academics and regulators}\label{ssec:pathwa}
Since the uptake of EnfTech produced by academia has been rather fortuitous and resembles a missed meeting of supply and demand, clear pathways for translating research outputs into deployable tools are urgently needed. These mechanisms should specify the safeguards required for legal compliance in enforcement activities, while providing practical guidance for implementation, considering organizational characteristics. Shared spaces, such as co-design processes and regulatory sandboxes where experimentation can safely and purposefully occur, can further reduce adoption barriers. Taking this step would ensure that research outputs translate into practical, impactful applications that would also address the cost requirement, which appears to be a paramount constraint.

Open-source software presents a promising avenue for adoption by regulatory authorities as underlined by some interviewees, primarily due to its accessibility and adaptability within existing enforcement workflows. In addition, the principles of Free and Open Source Software  and open data are well aligned with broader academic and policy objectives, including the promotion of open science and the necessity of transparency, cost cutting and independence in public administration \cite{Bouras2014}. 

Even if the code of some of the surveyed tools was released openly   \cite{mathur2019dark, kirkman2023darkdialogs, bouhoula2023automated}, our review identified several tools hosted on platforms like GitHub that lacked explicit licensing information. This omission poses a significant barrier to reuse and integration, as the absence of a clear license undermines legal clarity and restricts the potential for collaborative development and deployment in regulatory contexts or even just for inspiration. Thus, we urge DP tool-makers to consider using open-source models and releasing code accompanied by appropriate licenses.
%

Tools that are really needed in enforcement activities beyond DP detection may attract fewer academic citations than trending topics like LLMs and dark patterns, prompting reflection on whether researchers should prioritize tangible societal impact over academic performance metrics—or find ways to balance both. Existing academic inquiry in computing technology and DP aspires to have impact, but still falls short in practice to be a driver of useful innovation. However, research assessment increasingly recognizes diverse forms of research contribution beyond citation counts \cite{CoARA_2022} and the DP domain is a great example of how different stakeholders can work towards shared goals \cite{gray2024ontology,gray2024mobilizing}.
We need to create mechanisms for incentive alignment between these stakeholders, such as designing funding instruments tailored to joint academia-regulation objectives and encouraging R\&D business opportunities, for instance based on open-core licensing models where open source is combined with proprietary software features for specific clients. CP-P9 highlighted that a spin-off valued their partnership because it boosted its market visibility, underscoring the practical advantages of such cooperation.

Public venues for inter-sectoral dialogue, such as CNIL’s Researchers Days\footnote{\url{https://www.cnil.fr/en/tag/Research}}, the international Computers, Privacy and Data Protection Conference (CPDP)\footnote{\url{www.cpdpconferences.org/}}, and EU COST Action projects\footnote{\url{https://www.cost.eu/}}, can foster collaboration and maximize impact. National agencies like PEReN in France\footnote{\url{https://www.peren.gouv.fr/en/}} and international ones like the eLab of the European Commission \cite{Tuch_2023} offer tools to public administrations and thus represent potential partners. This cross-pollination would also help researchers achieve the societal impact increasingly demanded by funding agencies, ensuring that public investment serves its intended purpose, aligned with their institutional missions.

\subsection{Balance transparency and auditability with confidentiality}\label{ssec:balance}
Another challenge that needs to be overcome derives from misalignments between academic and regulatory goals and requirements. The results indicate that automation shows greater potential in early regulatory phases (such as Sweep collection and Screening), while later stages require expert human oversight.  Considerations for enabling accountability, including ways to ensure transparency and traceability of decisions must be embedded from the earliest stages of development. 

Researchers should document development processes to make adoption by public regulatory bodies more predictable when requesting resources, and decrease the risk of legal appeals which would undermine the tool’s intended purpose and shift the burden onto regulatory authorities. Computing technologies used for evidence collection must be auditable and explainable; if they rely on opaque AI systems (such as those based on LLMs, e.g., \cite{Deceptilens_2025,gundelach2023cookiescanner,shi202550}), they can only serve inspirational or exploratory functions, because they risk creating accountability gaps, an unbearable risk shared by all use of AI in legal and judicial systems \cite{Sartor2025}. LLM-based tools, in particular, require additional safeguards to address issues like hallucinations and inaccuracies,  considerations that we have only seen in \cite{Deceptilens_2025}.
The need for transparency and explainability aligns with international standards for accountability in science, including the HLEG guidelines on trustworthy AI \cite{Hleg_2019} and ethical research practices \cite{genAI_2024}, thus we encourage research in this direction. 

Even though transparency is key for scientific reproducibility and trustworthiness, visibility must be limited to prevent malicious actors from undermining enforcement efforts, such as developing countermeasures (e.g., hiding cookie banner code), and avoid exposing sensitive enforcement cases' information.
Solutions to reconcile these diverse requirements consist in adopting careful disclosure policies such as tiered access models (e.g., public summaries versus restricted technical details), using data augmentation and synthetic DP data (e.g., \cite{kocyigit2024augmenting}), publishing methodologies and open-source foundation models while protecting sensitive datasets and customized algorithms, using appropriate licenses, and explicitly agreeing on confidentiality terms.

\section{Conclusion}
\label{sec:conc}
This study draws on interviews with practitioners operating across the regulatory landscape of online deceptive patterns, with the aim of identifying how automated detection tools originating from academic research can be meaningfully integrated into Enforcement Technologies (EnfTech), to enhance the capacity of regulatory bodies in addressing manipulative online practices. Addressing the lack of enforcement with automated means may increase deterrence and contribute to a fairer digital market in the long run \cite{Riefa_Coll_2024}.

Our findings indicate that current academic tooling predominantly targets the detection of potential non-compliance based on interface-level features. However, this approach does not align with the evidentiary standards required in regulatory proceedings, where robust, legally admissible documentation must be gathered from the outset. Consequently, while academic tools may inspire the development of internal regulatory systems, they are rarely adopted directly. To bridge this gap, it is imperative to establish a shared terminology and promote the development of adaptable, openly accessible tools based on aligned incentives.

As deceptive patterns continue to evolve, it becomes increasingly urgent to identify points of convergence between research and regulatory practice. Future work will reverse the perspective, examining how automated tools might support the administration of justice, and exploring how legal professionals conceptualize and engage with deceptive patterns.

\begin{acks}
The authors wish to thank the interview participants for their input. We also wish to thank the organizers and attendees of the Lorentz Center workshop ``Fair patterns for online interfaces'' (Leiden, 29 Jan. - 2 Feb. 2024) for inspiring the discussions that lead to this work, and Gabriele Lenzini and the Interdisciplinary Centre for Security, Reliability and Trust (SnT) of the University of Luxembourg for supporting initial ideation. A. Rossi wishes to acknowledge the support of the Italian Ministry of University and Research for the project BRIEF “Biorobotics Research and Innovation Engineering Facilities” (IR0000036) funded under the National Recovery and Resilience Plan (NRRP), Mission 4 Component 2 Investment 3.1 funded by the European Union – NextGenerationEU, and the complementary actions to the NRRP “Fit4MedRob - Fit for Medical Robotics” Grant (PNC0000007).
\end{acks}

\bibliographystyle{ACM-Reference-Format}
\bibliography{refs.bib}

\appendix
\section{Interview Questions}
\label{app:questions}
\subsection{Part I -- The person and role}
\begin{enumerate}
\item Who are you and what do you do?
\item What is your role in the enforcement process?
\item does your work relate to a specific regulation / area of law?

-> What are you trying to enforce / stop / protect (and for who)?

-> What does 'evidence' look like for this work?

\end{enumerate}

\subsection{Interfaces and violations}
\begin{enumerate}
\item Now we want to focus on legal violations in interfaces:
\item Can you give us a couple of examples of these legal violations? (Are those coming to mind because you deal with them a lot or because they are most severe or other reasons?)
\end{enumerate}

\subsection{Tools to manage interfaces}
\begin{enumerate}
\item What kind of tools do you use? (may need to be broad on what a 'tool' is)

-> Who are the intended users and uses?

-> Are (any elements of) tools specifically for your own use / tailored, or commonly-accessible? If so, why?

\item Are there user / regulatory / other requirements that you consider when developing / selecting / using these tools?

-> Probe: constraints on time/resource/people / restrictions on what to do or how to do it, etc.

-> Probe: How do you set priorities? why?

-> Probe: e.g., hide the tools from businesses to avoid circumvention?

\item What evidence do you / your tools collect? what evidence it is not able to collect (and how do you collect it, if at all)?
\end{enumerate}

\subsection{Use of tools for evidence and process}
\begin{enumerate}
\item is your use of tools part of cases/instances or driven by types of evidence?
-> Are there recurring cases/instances, or challenges, that you see with interfaces?

\item Is there a process that tools fit into? What is it?
-> What do you do with the evidence?
-> How do you know that your tool is achieving its goal? How do you relate tools to regulations and law?

\item what change do you expect to see from the use of the (enforcement) tool(s)?
-> How do your actions relate to website interfaces / interface design(ers)?
 Are these tools robust enough to be able to be hold up in court?
\end{enumerate}

\subsection{Part II - present tools: views on the academic tools (probes)}
\begin{enumerate}
\item Are these approaches familiar to you?
\item If these are promising, where can they go next and how can they connect to your work?
\item If these are promising, would they already connect to your work, or do they need to address anything specific first?
\item If these are not promising, is there something which can be done to make them connect to your work? Or is it that you see a need for something else first? (prioritization issue)
\item If these are not promising, is it because you expect them to be used by some other actor for some other purpose? (i.e., connect to someone else's work and skills) (matching of support to the right actor)
\item If these are not promising, is that related to law or requirements for evidence?
\end{enumerate}

\subsection{Part III - Future work and Debrief}
\begin{enumerate}
\item Are there regulatory developments coming up which you want to respond to? (= future activities?)
\item Is there anything you would like to know about this research study?
\item Our goal - the usefulness of academic research and the tools it represents
\end{enumerate}

\newpage
\section{Interview Codebook}
\label{app:codebook}
\begin{table}[!htbp]
    \centering
    \small 
    \begin{tabular}{|l|l|}
    \hline
        \textbf{Code group} & \textbf{Code}\\
        \hline
        Role & \\
         & Type of organization\\
         & Expertise\\
         & Task\\
        \hline
        DP & \\
         & Example\\
         & Why it is a DP\\
         & Why it is not a DP\\
         & Definition\\
        \hline
        Evidence / Input data & \\
         & Example / type\\
        \hline
        Activities within investigation & \\
         & Data collection rules\\
         & Data scraping\\
         & Data analysis\\
         & Artificial user agent\\
         & AI | Manual\\
         & Flagging / filtering\\
         & Self-assessment\\
         & Derive altern./response action\\
         & Reporting outputs\\
         & ML tooling\\
         & Decision support\\
        \hline
        Development & \\
         & In-house\\
         & Cross-agency\\
         & External\\
        \hline
        Collaboration & \\
         & Partner\\
         & Governance\\
         & Motivation\\
        \hline
        Constraint / Requirement & \\
         & Cloud provider\\
         & Cost / efficiency / capacity\\
         & Sensitivity of data\\
         & Open source\\
         & Tech skills\\
         & Human check\\
         & Risks\\
         & Reliability / accuracy of results\\
         & Accountability of the process\\
         & Validity in judgment\\
         & Integration (tech+user)\\
         & General vs. case-specific\\
        \hline
        Future development & \\
        \hline
        Tool limitations & \\
         & Context\\
         & Too many factors\\
         & Only look at UI\\
         & Continuous DP evolution\\
        \hline
        Users & \\
         & Businesses\\
         & Regulators\\
         & Consumers / End-users\\
        \hline
        Regulations / Laws & \\
        \hline
    \end{tabular}
    \caption{The codebook which emerged from the analysis of the practitioner interviews.}
    \Description{Interview codebook. Code groups include Role, DP, Evidence / input data, Activities within investigation, Development, Collaboration, Constraint / Requirement, Future development, Tool limitations, Users, and Regulations / Laws.}
    \label{tab:codebook}
\end{table}

\clearpage
\onecolumn
\section{Overview of Tools}
\label{app:overview}
\begin{figure*}[ht]
    \centering
    \includegraphics[width=\textwidth]{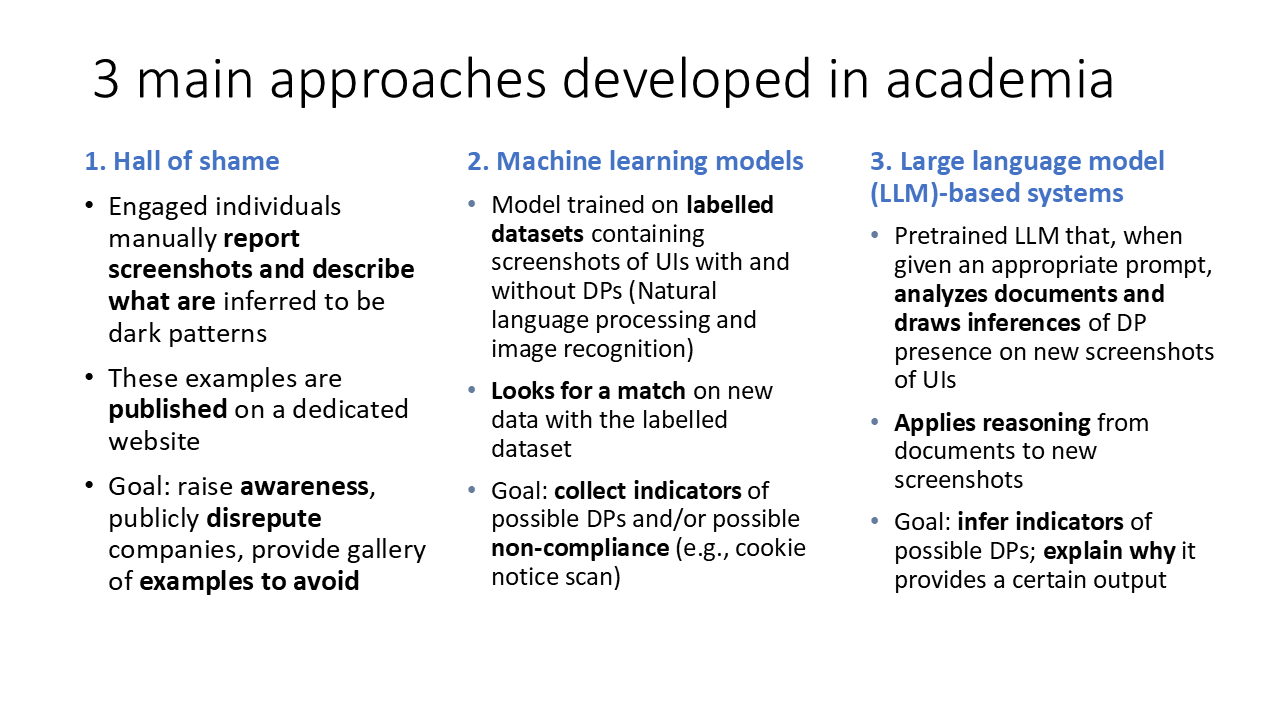}
    \caption{The overview of the three types of tools summarized to participants during the interviews: Hall of shame, Machine Learning (ML), and Large Language Model (LLM).}
    \label{fig:overview}
    \Description{The overview of the three types of tools summarized to participants during the interviews. These are Hall of shame, Machine learning models, and Large language model (LLM)-based systems.}
\end{figure*}

\end{document}